\documentclass[preprint,noshowpacs,noshowkeys,floatfix,aps,nofootinbib,pra]{revtex4-1}
\usepackage{graphicx}

\begin{document}

\title{Two--Dimensional Water in Graphene Nanocapillaries Simulated with Different Force Fields: Rhombic Versus Square Structures, Proton Ordering, and Phase Transitions }

\author{Shujuan Li}
\author{Burkhard Schmidt} 

\affiliation{
Institute for Mathematics, Freie Universit\"{a}t Berlin \\ Arnimallee 6, D-14195 Berlin, Germany}

\date{\today}

\begin{abstract}
Hydrogen bond patterns, proton ordering, and phase transitions of monolayer ice in two-dimensional hydrophobic confinement are fundamentally different from those found for bulk ice.
To investigate the behavior of quasi--2D ice, we perform molecular dynamics simulations of water confined between fixed graphene plates at a distance of 0.65 nm.
While experimental results are still limited and theoretical investigations are often based on a single force field model, this work presents a systematic study using different water force fields, i.~e. SPCE, TIP3P, TIP4P, TIP4P/ICE, TIP5P.
The water--graphene interaction is modeled by effective Lennard-Jones potentials previously derived from high--level ab initio CCSD(T) calculations of water adsorbed on graphene [Phys. Chem. Chem. Phys. \textbf{15}, 4995 (2013)].
The water occupancy of the graphene capillary at a pressure of 1000 MPa is determined to be between 13.5 and 13.9 water molecules per square nanometer, depending on the choice of the water force field.
Based on these densities, we explore the structure and dynamics of quasi--2D water for temperatures ranging from 200 K to about 600 K for each of the five force fields. 
To ensure complete sampling of the configurational space and to overcome barriers separating metastable structures, these simulations are based on the replica exchange molecular dynamics technique. 
We report different tetragonal hydrogen bond patterns which are classified as nearly square or as rhombic. 
While many of these arrangements are flat, in some cases puckered arrangements are found, too.
Also the proton ordering of the quasi-2D water structures is considered, allowing to identify them as ferroelectric, ferrielectric or antiferroelectric.
For temperatures between 200 K and 400 K we find several second--order phase transitions from one ice structure to another, changing in many cases both the arrangements of the oxygen atoms and the proton ordering.
For temperatures between 400 K and 600 K there are melting--like transitions from a monolayer of ice to a monolayer of liquid water.
These first--order phase transitions have a latent heat between 3.4 and 4.0 kJ/mol.
Both the values of the transition temperatures and of the latent heats display considerable model dependence for the five different water models investigated here. 
 
\end{abstract}

\maketitle
\section{Introduction}

Understanding the structure and phase behavior of water is of great interest due to its extraordinary properties and ubiquitous existence in our daily life. 
Under different conditions of pressure and temperature, bulk water can form numerous crystal structures, with the familiar ice Ih being just one of at least 17 crystalline phases \cite{Zheng_etal1991,Poole1992}. 
Less obvious are the low--dimensional ice--like structures of water confined in nanocapillaries or adsorbed at nanointerfaces which is a subject of considerable scientific interest due to important implications on biological systems, geological systems, and nanotechnological application \cite{Finney2004,Chandler2005,Bertrand2013,Carrasco2012,Gao2015}.
Low-dimensional water in nanoscale confinement exhibits profound differences both in structural and dynamic properties compared with bulk water and great progress has been made in understanding them.
While experimental studies are still rare, see e.~g. work on one dimensional confined water \cite{Kolesnikov2004,Kyakuno2016a,Kyakuno2011,Cambre2010,Wang2008,Maniwa2002,Kolesnikov2006,
Holt2006a,Byl2006a,Maniwa2005}, and two dimensional confined water \cite{He2012,Yang2009,Zheng2013,Kimmel2009,Algara2015}, there is an extensive body of theoretical investigations.
One part of these simulation studies is devoted to quasi--1D water confined in low--diameter nanotubes or other nanopores \cite{Hummer2001,Xu2011,Klameth2013,Berezhkovskii2002,Dellago2003,Kofinger2008,
Wang2009a,Noon2002,Wang2004a,Bai2006,Takaiwa2008,Alexiadis2008d,Bai2003,Koga2000,
Koga2001b,Li2015}, while another part is concerned with quasi--2D water locked between two parallel plates comprising of graphene or other materials \cite{Zangi2004a,Zangi2004,Johnston2010,Zangi2003b,Zangi2003d,Kaneko2014,DelosSantos2012,Strekalova2011,
Mazza2012,Han2010,Bai2012,Giovambattista2009,Bai2003a,Krott2013,Koga1997,Giovambattista2006,Kumar2005,
Meyer1999,Kumar2007,Mosaddeghi2012a,Corsetti2016,Zhao2015,Koga2000b,Koga2005,Werder2003,Zhu2015a,Zhao2014b,Cicero2008a,Ferguson2012a,
Kaneko2013,Qiu2013,Kaneko2014,Strauss2014,Zhao2014a,Mario2015a,Corsetti2005b,Chen2016,Corsetti2016}.
 
This work presents a computational study of the structure and dynamics of monolayer (quasi-2D) water confined between two parallel graphene sheets, a prototypical model system for hydrophobic confinement. 
It is motivated by a recent high-resolution electron miscroscopy imaging study by Algara-Siller et al. who found a strictly square ice lattice which is also supported by accompanying molecular dynamics (MD) simulations~\cite{Algara2015}. 
Subsequently, the existence of nanoconfined 'square ice' at room temperature was confirmed both by density functional theory (DFT) calculations~\cite{Chen2016,Corsetti2005b} and conventional MD simulations~\cite{Zhu2015a}.
However, in other studies of the lattice structure of quasi-2D ice alternative structures were found, e.~g. flat nearly square~\cite{Koga2005}, flat rhombic~\cite{Zhao2014a}, puckered rhombic~\cite{Zangi2003b}, puckered square \cite{Koga2005,Zhu2015a}, flat hexagonal \cite{Ferguson2012a}, puckered nearly square \cite{Bai2010a}, and even  Archimedean $4 \cdot 8^2$ tiling  structures have been reported~\cite{Zhao2014a}.

In addition to the determination of the above structures of the oxygen ions, also the question of proton ordering or the orientation of the permanent dipole moments of the water molecules is a highly interesting topic. 
For bulk water the concept of ferroelectricity is still elusive. 
While already predicted by Pauling in the 1930s \cite{Pauling1935}, there is no experimental evidence yet for the existence of ferroelectric ice XI under natural conditions on earth.
However, for ice wires and ice nanotubes in the quasi-1D confinement of low-diameter carbon nanotubes (CNTs), various ferroelectric, ferrielectric and anti-ferroelectric \cite{Kofinger2008,Nakamura2012a,Li2015} arrangements of water molecules could be identified in the last few years, however, only in simulations.
The present work deals with the question whether nanoconfined quasi--2D ice can also be ferroelectric for which there is indeed (limited) experimental evidence.
Thin ferroelectric ice layers can be grown on platinum surfaces \cite{Su1998,Iedema1998b} or can be found in hydration shells surrounding proteins \cite{LeBard2010}.
Also an atomic force microscopy imaging study probably suggests the possibility of ferroelectric water monolayers adsorbed on mica surfaces \cite{Spagnoli2003a}. 
However, there is no experimental evidence yet for ferroelectricity of water monolayers sandwiched between graphene plates.
So far, ferroelectric ordering in such systems has only been reported in MD simulation studies~\cite{Zangi2003b,Zhao2014a}. 

Once the large number of new topologies of monolayer ice confined in nanocapillaries has been explored,
another main aspect of the present work is related to the phase behaviour of quasi--2D water. 
Which of the above--mentioned water structures prevails at which temperature, and how can we characterize the phase transitions behavior between them?
It is expected that the melting and freezing behavior of nanoconfined water will be qualitatively different from that of bulk water.
For example, this has been shown for quasi--1D water in low-diameter CNTs \cite{Koga2001b,Takaiwa2008,Li2015} and also for quasi--2D water confined inside nanocapillaries.
For the latter case, there are different computational studies on the effect of different thermodynamic variables and for different confining surfaces \cite{Xu2011,Han2010,Koga2000b,Kaneko2014,Giovambattista2009,Chen2016,Koga2005,Zhu2015a}.
However, the effects of finite temperature on monolayer water confined between graphene sheets, in particular the nature of the underlying phase transitions, are yet to be comprehensively studied in the present work.

The present work aims at a systematic study of structures of quasi--2D nanoconfined water, including proton ordering, and the corresponding phase behavior.
Even though, as detailed above, there is already a substantial body of literature, a direct comparison is often not straightforward due to two different reasons.
First, different values of the underlying thermodynamic parameters, such as temperature, pressure and slit width were used.
Second, the simulations were often based on different force fields, both for the water--graphene and the water--water interaction.
Hence, a main focus of our study will be on the effect of different force fields.
For the water--graphene force field, we will use parametrizations based on high--level quantum chemical calculations, as we already did in our previous simulation studies of water in CNTs~\cite{Perez-Hernandez2013,Li2015,Lei2016}.
For the water--water force field, we will be using five standard water models, i.~e. SPCE \cite{Berendsen1987a}, TIP3P \cite{Jorgensen1983c}, TIP4P \cite{Jorgensen1983c,Jorgensen1985a}, TIP4P/ICE \cite{Abascal2005} and TIP5P \cite{Mahoney2000}.
Moreover, note that all studies presented here are for water confined in between parallel graphene sheets at a distance 0.65 nm.

Finally, we mention another aspect not covered sufficiently in most of the literature on quasi--2D water.
The coexistence of largely different minimum energy structures with very similar energies but very different water orientational properties (e.~g. ferroelectric, ferrielectric water and antiferroelectric water structures) presents a major challenge to finite temperature MD simulations of structure and dynamics of the nanoconfined ice. 
This is because the various water structures are highly metastable, with high barriers that typically cannot be overcome on a nanosecond timescale in conventional MD simulations for e.~g. $T=300$ K.  
Hence, in order to obtain a valid sampling of configurational space, we resort here to replica-exchange molecular dynamics (REMD)  \cite{Swendsen1986,Hansmann1997,Sugita1999,Trebst2006} simulation techniques which are based on swapping between different temperatures simulated in parallel.

\section{Simulation Methods}

\subsection{Force Fields}
\label{l-j}
The water--carbon interaction is modeled by pairwise additive Lennard-Jones (LJ) potential energy functions
\begin{equation}
U=4\sum_{ij}\epsilon_{ij} \left[\left(\frac{\sigma_{ij}}{r_{ij}}\right)^{12}-\left(\frac{\sigma_{ij}}{r_{ij}}\right)^6\right]
\end{equation}
where the attractive part varies as $r_{ij}^{-6}$ and the repulsive part varies as $r_{ij}^{-12}$ and where the summation extends over all atoms $(i)$ of the water molecules and all atoms $(j)$ in the graphene slab. 
The two sets of adjustable parameters are chosen as follows: 
the collision diameters $ \sigma_{ij} $ are deduced from standard vdW radii:  $\sigma_{CO}  =0.3157$ nm, $\sigma_{CH}=0.2726$ nm. 
The well depth parameters for water--carbon were parametrized previously by fitting to CCSD(T) high level quantum calculations for the water--graphene interaction \cite{Voloshina2011b,Perez-Hernandez2013}. 
In those works, the overall interaction strength $(\eta)$ and the dimensionless anisotropy parameter $(\delta)$ are defined as follows: 
\begin{equation}
\eta=\epsilon_{CO}+2\epsilon_{CH}, \quad \delta=1-(\epsilon_{CO}-2\epsilon_{CH})/\eta
\end{equation}
A relatively strong water--carbon interaction, $\eta=$1 kJ/mol, and a notable anisotropy parameter  $\delta=1$ between water and carbon were obtained in the mentioned publications. 
Those results were also confirmed by our subsequent work where the overall water--carbon interaction strength and anisotropy were obtained from fitting to DF-CCSD(T) results for water interacting with CNTs \cite{Lei2016}. 
 
For the water--water interaction, we use five different water models, namely the three--particle models SPCE and TIP3P, the four--particle models TIP4P and its variant TIP4P/ICE, and the five--particle model TIP5P.
\subsection{MD Simulations of Water Filling a Nanocapillary}
\label{MD}

\begin{figure}
\centering
\includegraphics[width=0.6\textwidth]{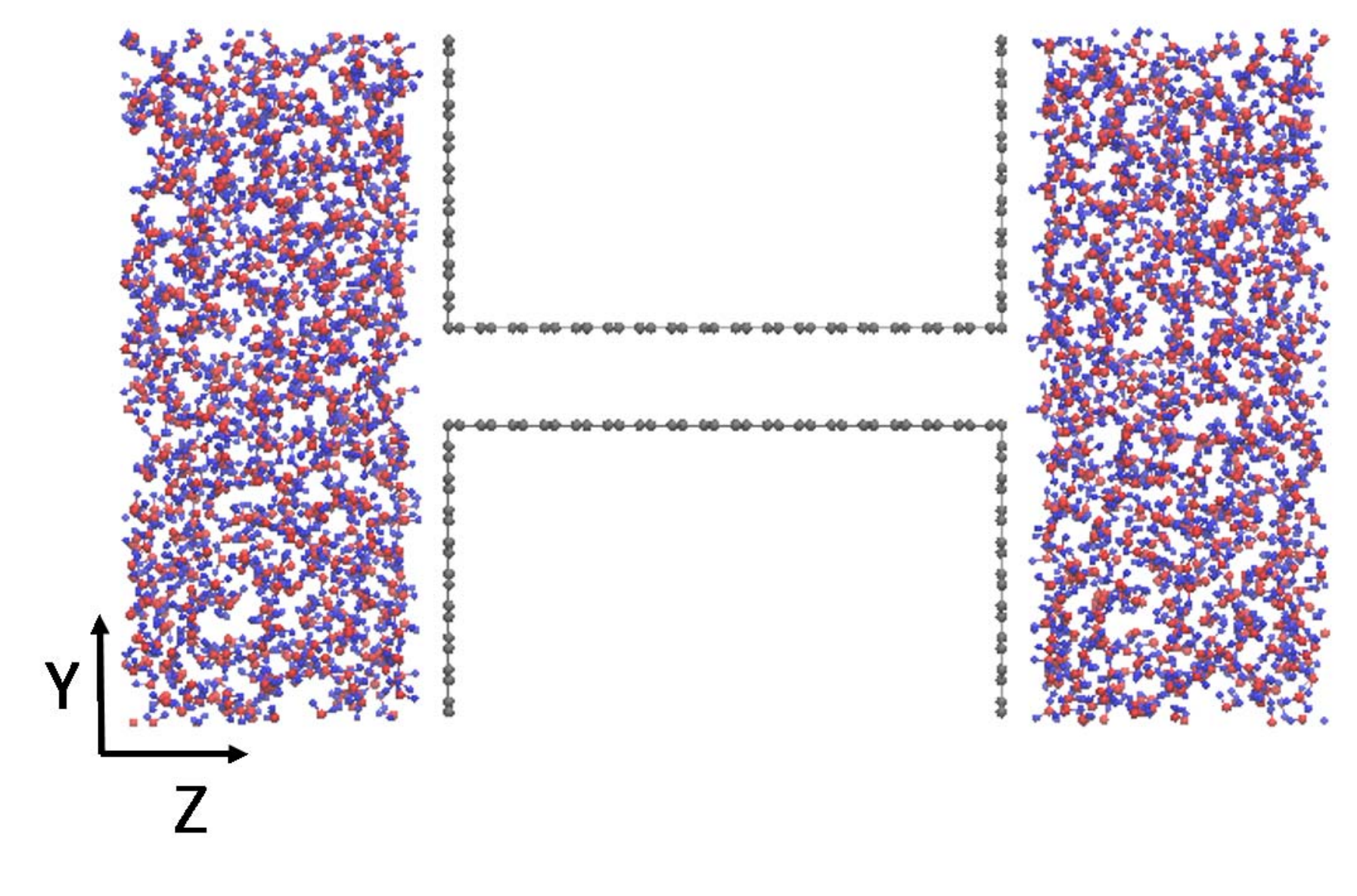}
\caption{Configuration for MD filling simulations to obtain the water densities confined inside graphene nanocapillaries. The MD simulation system initially contains 1000 molecules on each side.}
\label{Configuration}
\end{figure}

The number of water molecules, $N_{W}$, spontaneously entering a parallel graphene slit is obtained from molecular dynamics (MD) simulations for given temperature and pressure. 
Our MD simulation system contains 2000 molecules in the two sides of a simulation box separated by two graphene walls, see Fig.~\ref{Configuration}.  
The cross section of the simulation box is given by $l_x=3.689$ nm, and $l_y=4.626$ nm and its length is initially $l_z=8.032$ nm. 
The parts are connected by a nanocapillary which consists of two parallel graphene sheets the length and width of which are 3.689 nm and 3.692 nm. 
The distance $h$ between the graphene planes is 0.65 nm allowing one mono-layer of water to be accommodated within the capillary. 
The MD-simulations were carried out using the GROMACS 5.0.2 software package  \cite{Hess2008} within the $NPT$ ensemble. 
The water--water interaction is modeled in terms of five different water models, and
the water molecules and the graphene are assumed  to interact through the LJ potential energy functions introduced in Sec.~\ref{l-j}. 
The graphene walls as well as the nanocapillary are kept frozen during these simulations and the internal coordinates of the water molecules are constrained by the SETTLE algorithm \cite{Miyamoto1992}. 

During the initial equilibration period of 5 ns, the desired temperature, $T=300$ K, is controlled by the velocity-rescaling thermostat with a coupling constant of  $\tau=0.2$ ps \cite{Bussi2007} and the pressure, $P=0.1$ MPa or $P=1000$ MPa, is controlled by the pressure coupling Berendsen barostat \cite{Berendsen1984} acting along the $z$ direction.
This choice of thermostat/barostat provides an efficient means to reach a stable state at the beginning of a run.  
During the subsequent production run of another 5 ns length, the temperature is controlled by the Nos\'e-Hoover thermostat with a coupling constant $\tau=0.2$ ps \cite{Nose1984,Hoover1985} and the pressure is controlled by the pressure coupling Parrinello-Rahman barostat \cite{Parrinello1981,Jorgensen1983} which are known to yield more stable $NPT$ conditions than the velocity-rescaling thermostat and the Berendsen barostat, respectively.
The equations of motion are integrated using the leap-frog algorithm with a timestep of 1 fs with periodic boundary conditions in all directions.
For the LJ part of the water--water and the water--carbon interaction a cutoff radius of 0.9 nm is applied.
The Coulombic interaction of the water partial charges is treated by a real-space cutoff at 0.9 nm and the reciprocal part is treated by the Particle--Mesh Ewald (PME) method \cite{Darden1993a,Essmann1995}. 

In some of the simulations presented in Sec. \ref{thermal}, it was necessary to determine the (solid or liquid) phase of the quasi-2D ice confined in the nanocapillary.
We analyzed the mobility of the water molecules by calculating the mean square displacement (MSD), i.~e.,
\begin{equation}
\langle r^{2}(t)\rangle= \frac{1}{N_W}\sum_{i=0}^{N_W}\left(r_{i}(t)-r_{i}(0)\right)^2
\end{equation}
Here, $N_W$ is the number of the water molecules, and $r_{i}(t)-r_{i}(0)$ is the average distance they  travel in a given time $t$, here three nanoseconds.
If the MSD grows linearly with time, its slope can be related to the self-diffusion constant $D$ through the Einstein relation:  
\begin{equation}
D=\frac{1}{2d}\frac{\partial}{\partial t}\langle r^{2}(t)\rangle
\label{diffusion}
\end{equation}
where $d$ stands for the number of spatial dimensions, in this case two for the in-plane diffusion.

\subsection{REMD Simulations of Confined Water}
\label{REMD}
Replica exchange molecular dynamics (REMD) simulations are performed to study the structure and dynamics of water confined inside a 0.65 nm wide graphene slit, now without the surrounding water reservoirs.
These simulations are carried out within the $NVT$ ensemble, with constant number of water molecules in the capillary, $N_{W}$, taken from the MD filling simulations as described above.
The force field between water and graphene is the same as in the filling simulations, see also Sec.~\ref{l-j}.
Again, we use the five different water models listed there. 
The temperature, $T=300$ K, is controlled by the velocity-rescaling thermostat. 
The size of the nanocapillary area is 5.964 nm $\times$ 5.658 nm.
 
\begin{figure}
\centering
\includegraphics[width=0.8\textwidth]{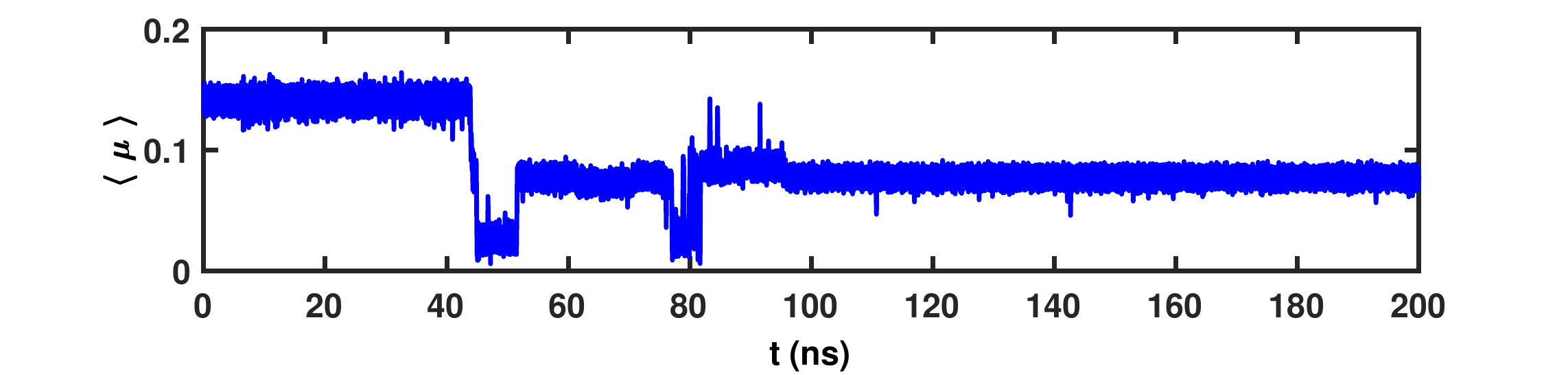}
\caption{Time evolution of dimensionless polarization for 467 water molecules inside a graphene nanocapillary for $T=$300 K. Note the metastable phases extending over tens of nanoseconds.}
\label{spce_mtot_200ns}
\end{figure}

When using conventional MD simulation techniques there are sampling problems connected with the rare events of transitions between metastable structures of water confined in the graphene slit. 
This is illustrated here for the dimensionless polarization $\langle \mu \rangle$ which is defined as the sum of the water dipoles projected onto the graphene planes, divided by the number of water molecules, $N_{W}$, and by the dipole moment, $\mu_0$, of a single water molecule which is 2.35 D \cite{Kusalik1994a}, 2.35 D \cite{Jorgensen1983,Mahoney2000}, 2.18 D \cite{Jorgensen1983,Kusalik1994a,Mahoney2000}, 2.43 D  \cite{Abascal2005} and 2.29 D \cite{Mahoney2000} for the SPCE, TIP3P, TIP4P, TIP4P/Ice, and TIP5P models, respectively. 
\begin{equation}
\langle \mu \rangle=\frac{1}{N_W \mu_0} \sum_{i=1}^{N_W} |\mu_i| 
\end{equation}
where ${\mu_i}$ is the dipole moment for the $i$-th water molecule along the graphene planes. 
As an example, Fig.~\ref{spce_mtot_200ns} shows the time evolution of the dimensionless polarization $\langle \mu \rangle$ for 467 water molecules inside a graphene slit for $T=300$ K simulated with the SPCE water model. 
Even though the simulation period is rather long (200 ns), we observe only very few transition events between various metastable water states for this trajectory. 
Obviously, the reason why such events are so rare is because they involve concerted rotations of many (or even all!) water molecules. 
Hence, it is computationally too expensive to sample the whole phase space of the system with conventional MD simulations. 
Instead, in the present work we make use of the REMD technique which is based on an ensemble of non-interacting MD simulations for different temperatures. 
In analogy to conventional Metropolis Monte Carlo simulations building on random walks in configuration space, the REMD algorithm represents a random walk in temperature space. 
The motivation is that broader sampling can be obtained at high temperatures, from where the configurations can swap to the lower temperatures. 
Thus, the simulated systems can overcome barriers between local minima of the energy through exchanging configurations between two neighboring temperatures \cite{Swendsen1986, Hansmann1997, Sugita1999}. 
The exchange between temperatures $i$ and $j$ is governed by a Metropolis-Hasting algorithm  \cite{Metropolis1953}, which satisfies the detailed balance condition \cite{Sugita1999}.
The resulting exchange probability is given by
\begin{equation}
P_{ij}=\min\{1,\exp[(\beta _i -\beta _j)(U(r_{i})-U(r_{j})]\}
\end{equation}
where $U(r_{i})$ and $U(r_{j})$ specify the potential energy of the configurations for the two temperatures and $\beta_{i,j}=\frac{1}{k_B T_{i,j}}$  is the inverse temperature and $k_B$ is the Boltzmann constant. 

Our REMD simulations are performed using MPI GROMACS 5.0 \cite{Hess2008} in an $NVT$ ensemble. 
The $N_T$ temperatures are distributed exponentially according to 
\begin{equation}
T_i=T_0 e^{k i}, \quad 0 \leq i \leq N_T
\end{equation}
where the temperatures range between $T_{min}=T_0$ and $T_{max}=T_0e^{kN_T}$ and where the parameter $k$ can be tuned to obtain temperature intervals allowing for sufficient acceptance probabilities $P_{ij}$ which should be typically within $0.2\ldots 0.3$ \cite{Hess2008}.
In some cases, however, it proved necessary to manually adjust the temperatures to meet this requirement, see Tab. \ref {Temp}. 
In our simulations of water confined in a graphene nanocapillary, the temperature distributions are ranging from $T_{min}=200$ K, where all the water structures are practically frozen, to different $T_{max} \approx 600$ K  where replicas are not trapped in local energy minima anymore.
The number of temperatures, $N_T$, which is 28 for SPCE and 30 for the other four water models, depends on the acceptance probabilities.

In practice, the REMD scheme is initialized by running conventional MD simulations of 1 ns length, to achieve equilibration for each of the temperatures separately.
Then short REMD simulations (100 ps) were carried out to validate the acceptance probability between adjacent replicas and/or to calibrate the above parameter $k$ where exchanges are attempted every 1 ps.
Afterwards, long REMD simulations (with a total length of 20 ns) are performed which are the basis of our analyis given in Sec. \ref{thermal}.

\section{Results and Discussion}

\subsection{Determination of Graphene Nanocapillary Water Filling}

Before investigating the structure and dynamics of confined water, we first have to determine the water occupancy of the graphene capillary, i.~e. the number of water molecules entering spontaneously the graphene slit which connects the two water reservoirs containing 1000 water molecules each as shown in Fig. \ref{Configuration}.
Similar filling studies can be found in Refs.~\cite{Algara2015,Zhu2015a} but there a systematic investigation of the effects of different (water--water and water--carbon) force fields, such as reported in Ref.~\cite{Li2015} for CNTs, is still missing.
In this part of our investigation, MD simulations using the $NPT$ approach of Sec.~\ref{MD}
are applied to investigate the spontaneous filling process and determine the number of water molecules $N_{W}$ (per square nanometer 1 nm$^2$) for different temperatures and different pressures. 
In order to study the influence of the force fields, three series of simulations are performed.

\begin{figure}
\centering
\includegraphics[width=0.5\textwidth]{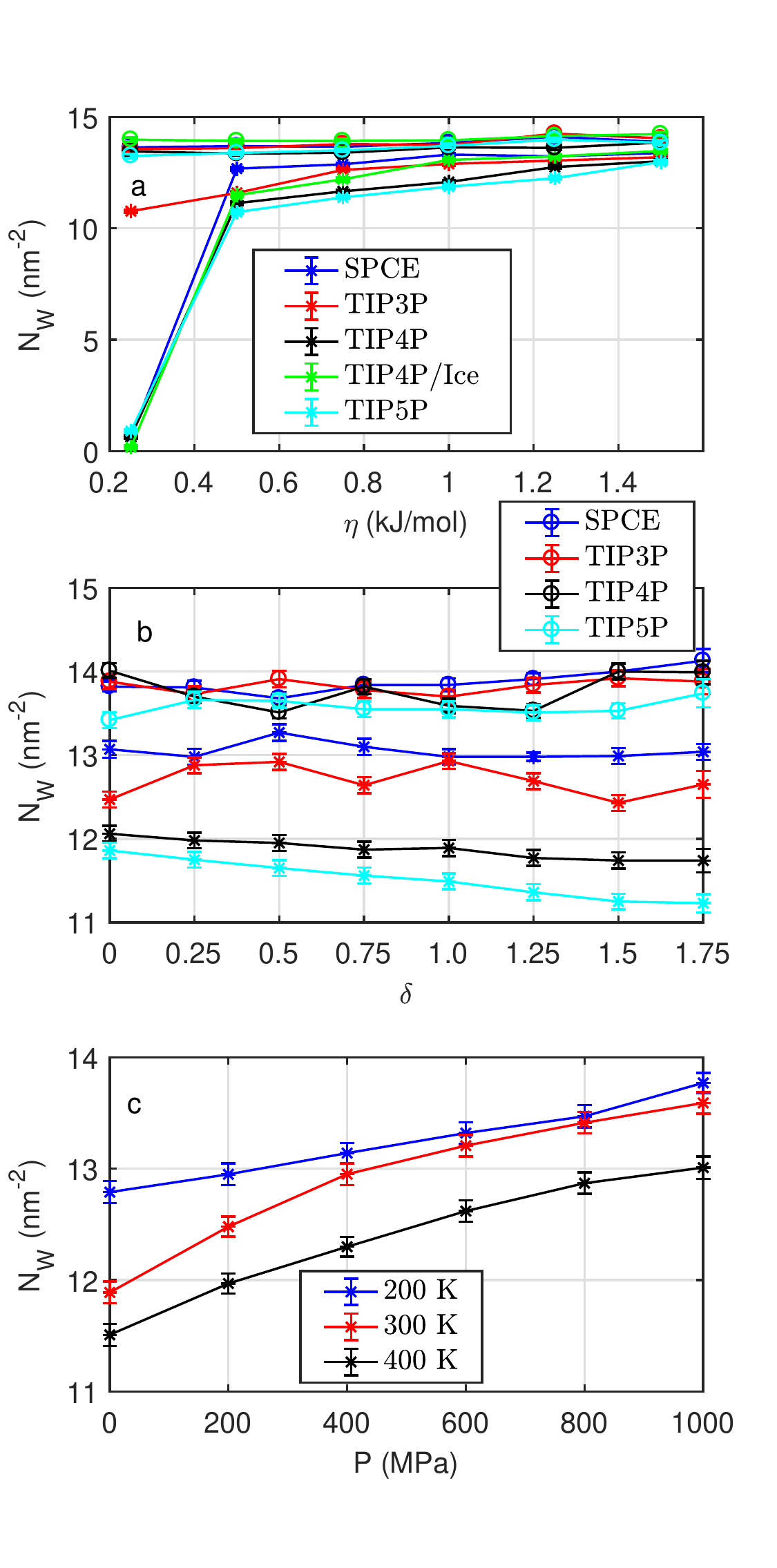}
\caption{(a) Influence of water models on water occupancy, $N_W$, of a graphene nanocapillary, as a function of the water--carbon interaction strength $\eta$, for isotropic interaction, $\delta = 0$,  $T=300$ K and two different pressure 0.1 MPa (stars) or $P=1000$ MPa (circles). (b) Influence of anisotropy parameter $\delta$ on water occupancy, $N_W$, for fixed interaction strength, $\eta=1$ kJ/mol, for $T=300$ K, and $P=0.1$ MPa (stars) or $P=1000$ MPa (circles). (c) Influence of pressure and temperature on water occupancy, $N_W$, for fixed anisotropy, $\delta=1$, $\eta=1$  kJ/mol, and for TIP4P water model.}
\label{density}
\end{figure}

First, the effect of the water--carbon interaction strength $\eta$ on the water occupancy is investigated for different water--water interaction models. 
Our results for isotropic water--carbon interaction ($\delta=0$) are shown in Fig.~\ref{density} a, for $T=300$ K and for two different pressures, 0.1 MPa and 1000 MPa.
By varying the interaction strength $\eta$ between 0.25 kJ/mol and 1.5 kJ/mol, we simulate the transition from hydrophobic to hydrophilic graphene for the water models SPCE, TIP3P, TIP4P, TIP4P/ICE, and TIP5P. 
For ambient pressure, $P = 0.1$ MPa, water is repelled from the interior of the graphene nanocapillary below a certain value of $\eta$. 
Interestingly, that threshold appears to be similar ($\eta \approx 0.5$ kJ/mol) for four out of five water models. 
Only for the TIP3P model, water spontaneously fills the graphene slit for $P=0.1$ MPa already for $\eta=0.25$ kJ/mol.
Above the respective threshold values, the water filling quickly rises and reaches saturation.
For high pressure (1000 MPa), water can fill the graphene slit practically without barrier, independent of the interaction strength, but the density is only slightly higher than for ambient pressure 0.1 MPa. 

Second, the effect of the anisotropy $\delta$ of the water--carbon interaction on the water occupancy is investigated, again for different water models. 
In contrast to the effect of the interaction strength $\eta$, the anisotropy $\delta$ does not affect $N_{W}$ notably, as shown in Fig.~\ref{density} (b) for ambient pressure $P=0.1$ MPa and high pressure $P=1000$ MPa. 
In contrast to bulk water at ambient conditions, where the difference in the water density simulated with different water models is negligible \cite{Vega2005}, this is not the case for our results shown in Fig.~\ref{density} (a) and Fig.~\ref{density} (b), where the water occupancy  reaches notably different values.
Hence, this difference can be considered as an effect of quasi--2D confinement in the graphene slit.
On the contrary, for $P=1000$ MPa, the water densities display much less differences between the different water models. 

Finally, the effect of pressure $P$ on the water occupancy is shown in panel (c) of Fig.~\ref{density}, in this case for the TIP4P water model only. 
The values $N_W$ increase nearly linearly with the pressure for $T = 200$ K and $T = 400$ K, where as for $T = 300$ K the increase is first fast and then slows down. 
Overall, the number $N_W$ for low temperature is higher than for high temperature, which can be caused by the different structures adopted at different temperatures, see below.

In summary, the above calculations show that the water occupancy of a graphene slit reached by spontaneous filling depends much more sensitively on the choice of the water model for ambient conditions than for high pressure (1000 MPa). 
In the former case, there is a threshold with the effect of interaction strength $\eta$ at ambient conditions for all investigated water models but not for TIP3P. 
Throughout the remainder of this work, a value of $\eta=1$ kJ/mol and $\delta =1$ will be used which is in agreement with our previous CCSD studies \cite{Lei2016} and also with our previous simulations of water confined inside CNTs \cite{Li2015}.  
The resulting water densities found for the five different water models at $P=1000$ MPa high pressure condition are listed in Tab.~\ref{water number}. 
These values will be used consistently both for the mimimum energy configurations and for the finite temperature REMD simulations described in the following two subsections.
\subsection{Minimum Energy Structures}
\label{ice}

This section is concerned with structure and polarization of water confined inside a graphene slit of 0.65 nm width within which water can form quasi-two-dimensional, single layer ice structures. 
First, we will discuss minimum energy structures which are obtained by means of a steepest decent algorithm implemented in the GROMACS 5.0 software package \cite{Hess2008}.
In order to sample the multitude of local minima of the high--dimensional potential energy landscape, a large number of minimizations were performed, which were initialized from snapshots of REMD trajectories (see Sec. \ref{REMD}), performed within the $NVT$ ensemble with periodic boundary conditions along the ice plane. 
In these calculations, the number of water molecules, $N_{W}$, is taken from the results of the filling simulations of a graphene nanocapillary with an area of 33.74 nm$^2$ under 1000 MPa, as discussed above.
By appropriate scaling of the water occupancy summarized in Tab. \ref{water number}, we obtained  $N_{W}=467, 462, 458, 467, 457$ for SPCE, TIP3P, TIP4P, TIP4P/ICE and TIP5P, respectively.

The H-bond networks of the quasi-2D water minimum energy structures are characterized in Fig.~\ref{min_E_structures} and Tab.~\ref{INT_table} where we use the following abbreviations: F, flat; P, puckered; S, square; R, rhombic; and N, nearly. 
The classification of the tetragons is mainly based on the distributions of the oxygen-oxygen-oxygen angles, $\alpha$, defined between the nearest neighboring water molecules, see Fig.~\ref{angle}. 
Depending on the tilt angle, $\tau$, we distinguish between nearly square (NS) ($|\tau|\leq 5^\circ$) and rhombic (R) else.

The proton ordering of the confined water can be quantified on the basis of the dimensionless polarization $\langle \mu \rangle$ introduced in Sec.~\ref{REMD}.
Here we classify minimum energy structures as ferroelectric (FE) for $0.9 \le \langle \mu \rangle < 1$, ferrielectric water (FI) for $0.1 \le \langle \mu \rangle <0.9$ and antiferroelectric (AF) with $\langle \mu \rangle< 0.1$.

The water--water and water--carbon interaction energies are also given in Tab.~\ref{INT_table}.
Note that in the following only the most stable, ordered minimum energy structures are discussed for each of the five water models under consideration. 

\subsubsection{SPCE Water Model}
\begin{figure}
\centering
\includegraphics[width=1.0\textwidth]{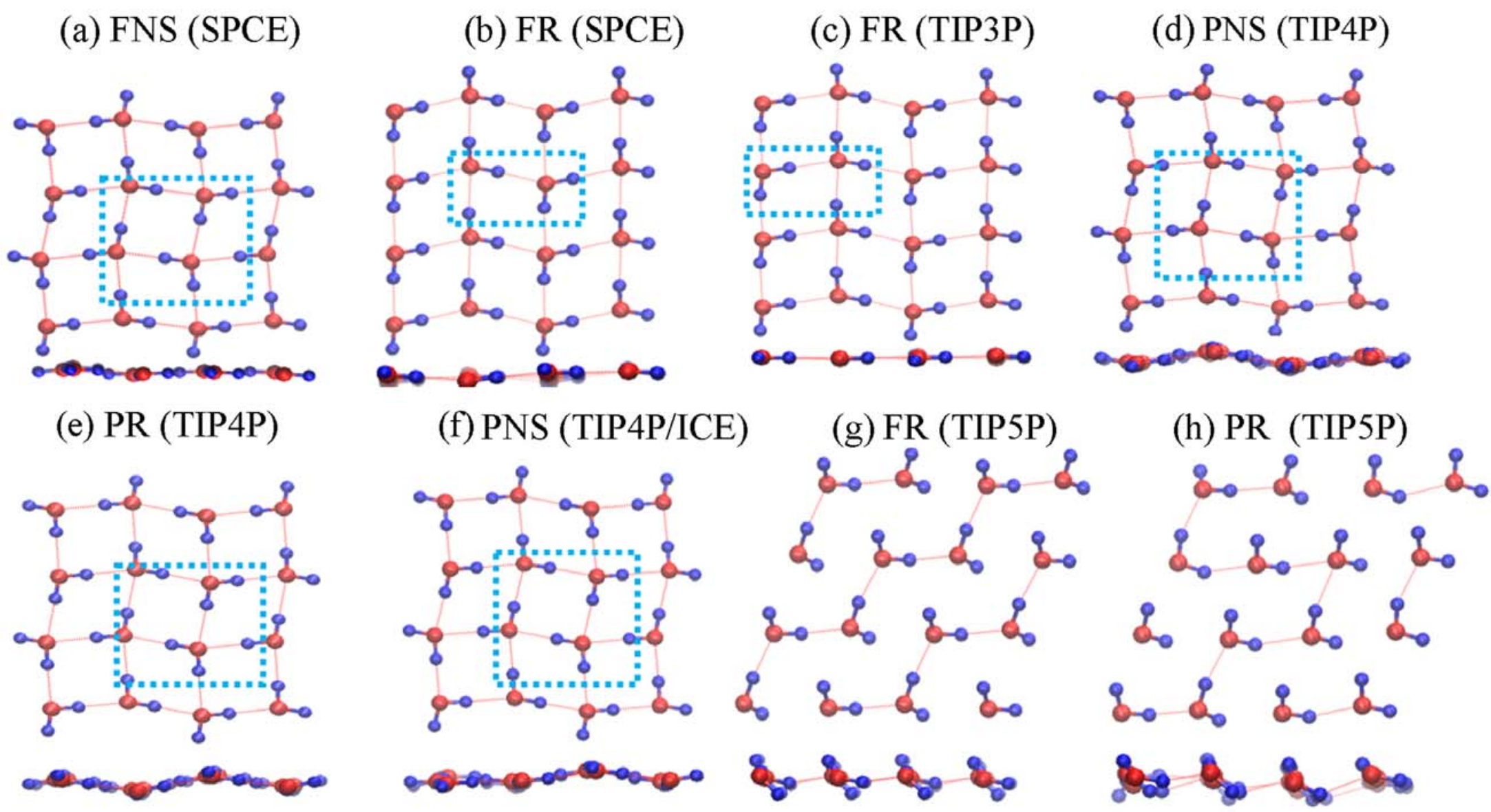}
\caption{Minimum energy structures of water confined inside a graphene nanocapillary for various water models indicated in brackets. Here, the blue rectangles are unit cells. The structures are denoted as follows: F = flat; P = puckered; N = nearly; S= square and R = rhombic, respectively. Note that these structures are also characterized in Fig. \ref{angle} and in Tab. \ref{INT_table}}
\label{min_E_structures}
\end{figure}

\begin{figure}
\centering
\includegraphics[width=0.6\textwidth]{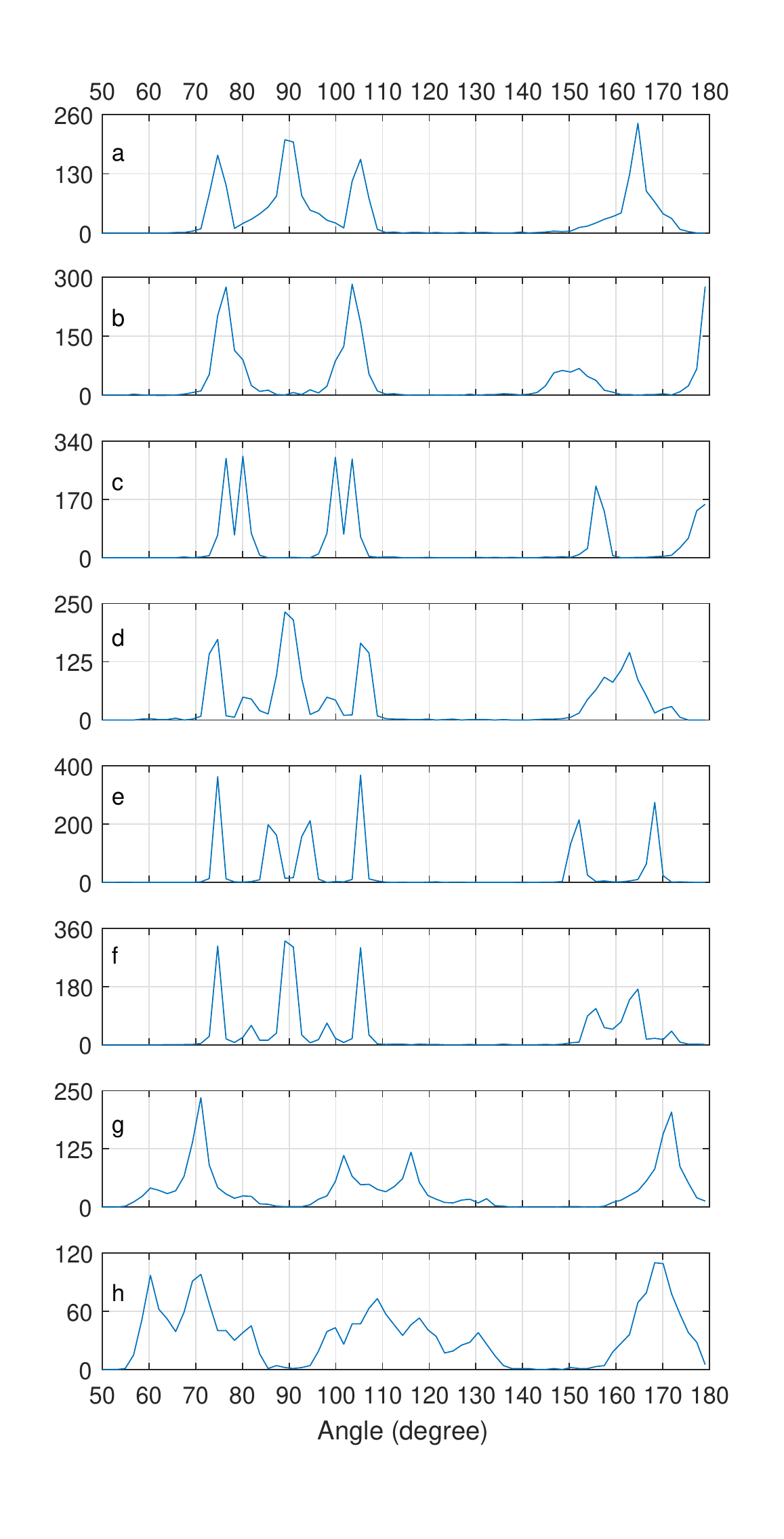}
\caption{Distribution of oxygen angles $\alpha$ of quasi--2D water confined inside a graphene nanocapillary. Distributions are obtained from minimum energy structures found for SPCE (a,b), TIP3P(c), TIP4P(d,e), TIP4P/ICE(f), and TIP5P(g,h) water model.}
\label{angle}
\end{figure}

The minimum energy structures for water inside a graphene slit simulated by the traditional three-site SPCE water model can be assigned to two different classes of monolayer ice, a flat nearly square (FNS) structure a
and a flat rhombic (FR) structure b.
The top view of Fig.~\ref{min_E_structures} shows that structure a (FNS) is almost square ice which can be also seen from the corresponding angle $\alpha$ distribution in Fig.~\ref{angle} where the peaks are centered at $90^{\circ}$, $75 ^{\circ}$ and $105^{\circ}$. 
Note that the strength of the $90 ^{\circ}$ peak matches the sum of the other two. 
Additionally, the peak at $165 ^{\circ}$ listed in Tab. \ref{INT_table} and shown in Fig.~\ref{angle} represents the slightly zigzag lines connecting the oxygen atoms. 
The end view of structure a (FNS) shows that the water molecules are nearly in one plane, i.e. the ice layer is indeed almost flat.

Structure b (FR) is flat rhombic with sharp angular peaks located at $\alpha =77^{\circ}$ and $103 ^{\circ}$ in Fig. \ref{angle}. 
Furthermore, the peaks at $152^{\circ}$ and $179^{\circ}$ indicate that there are zigzag (horizontal) lines and straight (vertical) lines in the network of the O atoms in Fig.~\ref{min_E_structures}. 
Finally, the end view of structure b (FR) shows that the water molecules are again nearly in the same plane.

While the networks of the O-atoms differ only slightly, the proton ordering of structures a and b is completely different, see also the average dipole moments $\langle \mu \rangle$ in the fourth column of Table~\ref{INT_table}. 
For structure a (FNS) we find a very low polarization $\langle\mu\rangle = 0.05$ thus rendering this structure AF. 
We can see that within each unit cell (light blue rectangle in Fig.~\ref{min_E_structures})  the four dipole vectors add up to nearly zero. 
For structure b (FR), however, the water dipoles point toward two different directions forming an angle of $82^\circ$, thus rendering this structure FI with polarization $\langle \mu \rangle$= 0.75, in agreement with $\cos 41^{\circ}=0.755$.  
While the water--water interaction energy $E_{W-W}$ is 0.65 kJ/mol (or 1.5 \%) lower for structure a (FNS) than for structure b (FR), the water--carbon interaction energies $E_{W-C}$ are essentially identical.  

\subsubsection{TIP3P Water Model}

For the TIP3P water model, another traditional three-site water force field, we found only one minimum energy ice structure, the flat rhombic (FR) structure c. 
In the top view of Fig.~\ref{min_E_structures} and angle distribution of Fig.~\ref{angle}, it can be seen that structure c consists of two different rhombic sub-structures.
Tab.~\ref{INT_table} reveals that they are distinguished by two sets of angles $\alpha$, centered at $77^{\circ}$ and $103^{\circ}$ versus $80^{\circ}$ and $100^{\circ}$.
Note that the former substructure is almost the same as structure b (FR) found for the SPCE water model.
Again, the angle peaks at $156 ^{\circ}$ and $179^{\circ}$  indicate that the network of O atoms can be characterized by (horizontal) zigzag oxygen lines and straight (vertical) oxygen lines, again similar to structure b (FR). 
Also the water molecules  of structure c (FR) are in one plane as displayed in the end view of Fig.~\ref{min_E_structures} (c). 

In analogy to structure b (FR), also structure c (FR) is found to be FI with a moderately high polarization $\langle \mu \rangle =0.74$.
Here the water dipoles are oriented along two different directions that form an angle of approximately $85^{\circ}$ with each other. 
 
\subsubsection{ TIP4P Water Model}

For the four-site TIP4P water model, we found two types of minimum energy ice structures. The puckered nearly square  (PNS) structure d with angle $\alpha$ peaked around $90^{\circ}$, $75^{\circ}$ and $105^{\circ}$ is very similar to structure a (FNS) found for the SPCE water model. 
Again, the strength of the $90^{\circ}$ peak matches the strength of the sum of the other two. 
In addition, the angle distribution of structure d displays minor peaks near $81^{\circ}$ and $99 ^{\circ}$. 
In both dimensions of the ice monolayer, the oxygen atoms are connected by zigzag lines with angles around $\alpha$=$163 ^{\circ}$. 
In addition, we also find a puckered rhombic (PR) minimum energy structure e.   
In the top view of Fig.~\ref{min_E_structures} and angle distribution of Fig.~\ref{angle} we can see that structure e consists of two different rhombic sub-structures distinguished by two sets of angular peaks, see also Tab.~\ref{INT_table}. 
The angular peaks at $152 ^{\circ}$ and $168 ^{\circ}$ indicate that there are zigzag oxygen lines in both dimensions, but with a different curvature. 
Even though the difference between structure d (PNS) and e (PR) appears to be not very pronounced in Fig.~\ref{min_E_structures}, the angle distribution in Fig.~\ref{angle} shows that almost half of the intensity for structure d (PNS) is found at $90 ^{\circ}$, whereas there is weak tilting ($85^{\circ}$, $95^{\circ}$) in structure e (PR). 
Note that both in structure d (PNS) and e (PR) the water layer is puckered, in marked contrast to the flat structures a, b and c observed in calculations for the SPCE and TIP3P water models. 
Obviously, this puckering is a consequence of moving the negative charge from the O--atom to a fourth potential site (dummy atom) located on the H--O--H bisector while keeping the LJ--term on the O--atom in the TIP4P model. 
Thus, slight puckering can reduce the LJ-repulsion while increasing the Coulomb attraction between nearest neighbor water molecules.

Finally, both structures d (PNS) and e (PR) are AF with polarization $\langle \mu \rangle$ near zero, hence in that respect very similar to structure a (FNS). 
The energy $E_{W-W}$ is 0.72 kJ/mol lower for structure d (PNS) than for structure e (PR). 

\subsubsection{TIP4P/ICE Water Model}

The TIP4P/ICE model is a modification of the original TIP4P model, aiming at an improved reproduction of the phase diagram of bulk water, but without a deterioration of the remaining bulk water properties.
In our work on quasi-2D ice, we found one puckered nearly square (PNS) configuration f for the water model TIP4P/ICE, which is quite similar to structure d (PNS) found for the original TIP4P water model, as shown in Figs.~\ref{min_E_structures} and \ref{angle}. 
Also the AF proton ordering of structure f (PNS) is quite similar to that of structure d (PNS).   
However, the oxygen network of the former one appears to be a bit less puckered. 

\subsubsection{TIP5P Water Model}
The five-site TIP5P water model was originally introduced to reproduce the bulk water density over a very wide range of pressures, including the density maximum near $T\approx 277$ K at ambient pressure.
When using this water model in simulations of quasi-2D water confined in a graphene nanocapillary, we found two minimum energy ice forms, namely the flat rhombic (FR) structure g and the puckered rhombic (PR) structure h. 
With the angular peaks located at $71^{\circ}$, $101^{\circ}$ and $116^{\circ}$, the FR structure g is regular but more tilted than the rhombic structures discussed above. 
Moreover, the peaks are not symmetric around $90^{\circ}$ for water model TIP5P.
Finally, there is another peak at $172 ^{\circ}$ which indicates a slight curvature of the O--atom connectors.

The puckered rhombic (PR) structure h does not present a single crystalline form like all structures mentioned thus far, but rather it contains different rhombic sub-structures. 
This is confirmed by the angle distribution in Fig.~\ref{angle}, displaying several groups of peaks around $60 ^{\circ}$ and around $115^{\circ}$. 
Again the peak at $169 ^{\circ}$ shows that there are zigzag oxygen lines in both dimensions similar to structure g (FR).

In contrast to our findings for the three-- and four--site water models, both in structures g and h some of the O--H$\ldots$O arrangements deviate substantially from linearity. 
Hence, H-bonds are not always drawn in Fig.~\ref{min_E_structures} g and h. 
The reason for this is the tetrahedral arrangement of the charges in the TIP5P model mimicking the lone pair electrons \cite{Agmon2012}. 
Another consequence is that most of the H atoms are located in two different planes above and below the plane spanned by the O-atoms, as indicated in the side views.
While in structure g (FR) the H--atoms are distributed equally between the two planes, the arrangement of the H--atoms of structure h (PR) appears to be more disordered. 
Nonetheless, as far as the proton ordering is concerned, both structures g (FR) and h (PR) are FE with very high polarizations of 0.97 and 0.94, respectively.

\subsection{Temperature Effects and Structural Transitions}
\label{thermal}

In this section we discuss the results of our REMD simulations within the $NVT$ setting for the quasi-2D water system confined between two graphene layers.
Special emphasis is on thermal effects and structural transitions within a temperature range between 200 K and about 600 K, for technical details see Sec.~\ref{REMD}.  
Where possible, we want to identify the 2D minimum energy structures introduced in Sec.~\ref{ice} and try to estimate at which temperatures they occur. 
In that context, interesting observations are the temperature dependence of the distributions of oxygen angles $\alpha$ shown in Fig.~\ref{Angle_surface} and the distributions of the (dimensionless) polarization $\langle \mu \rangle$ in Fig.~\ref{Mtot} allowing to identify FE, FI and AF arrangements of the protons. 
In addition, we analyze the caloric curves along with a decomposition of the averaged energies into kinetic energy, LJ (water--carbon and water--water) and Coulomb (water--water only) potential energies, see Fig.~\ref{energy}. 
We shall use the water--water potential energy to determine at which temperature structural transitions occur and, where possible, to determine the latent heat for 2D ice structural transitions. 
Moreover, the structural transitions can be classified according to the definition of phase transitions. 
In first--order transitions first derivatives of the energy undergo discontinuous changes. 
In second--order transitions first derivatives of the energy are continuous but the second derivatives are discontinuous.
In addition, we distinguish liquid from solid phases by analyzing the self-diffusion constant from Eq. (\ref{diffusion}), based on normal $NVT$ simulations for selected temperatures.

To further analyze the structure of water confined in a graphene nanocapillary, we also analyze the H-bonding networks  obtained from our REMD simulations as shown in Fig.~\ref{H-B}. 
To go beyond the number of H--bonds each water molecule is engaged in, the pattern of H--bonding between nearest neighbors is characterized here by joint probabilities $p_{n_a,n_d}$ of a water molecule to act $n_d$ times as a donor and $n_a$ times as an acceptor at the same time \cite{Agmon2012}. 
To account for the floppy arrangement of water molecules in our simulations, we use a relaxed criterion for the detection of H-bonds, i.~e. , O--O distance up to 0.35 nm and deviation from linearity of the O--H $\cdots$ O arrangement up to 45 degrees. 

\subsubsection{SPCE Water Model}

\begin{figure}
\centering
\includegraphics[width=0.7\textwidth]{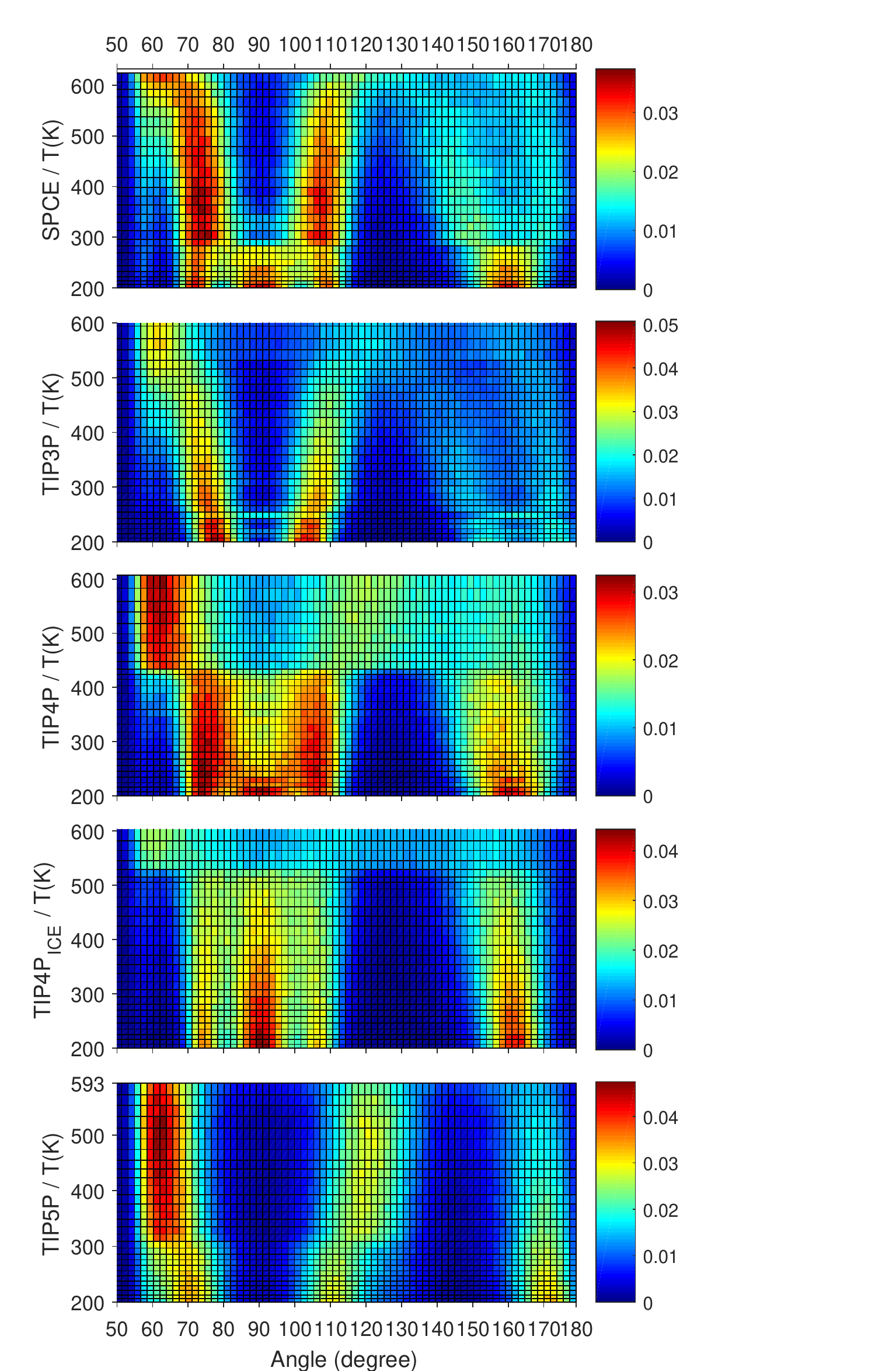}
\caption{Distribution of oxygen angles $\alpha$ for quasi--2D water confined inside a graphene nanocapillary as a function of temperature. Obtained from REMD simulations with $\eta=1$ kJ/mol, $\delta=1$, and for the SPCE, TIP3P, TIP4P, TIP4P/ICE, and TIP5P water models from top to bottom for a pressure of 1000 MPa.}
\label{Angle_surface}
\end{figure}

\begin{figure}
\centering
\includegraphics[width=1.0\textwidth]{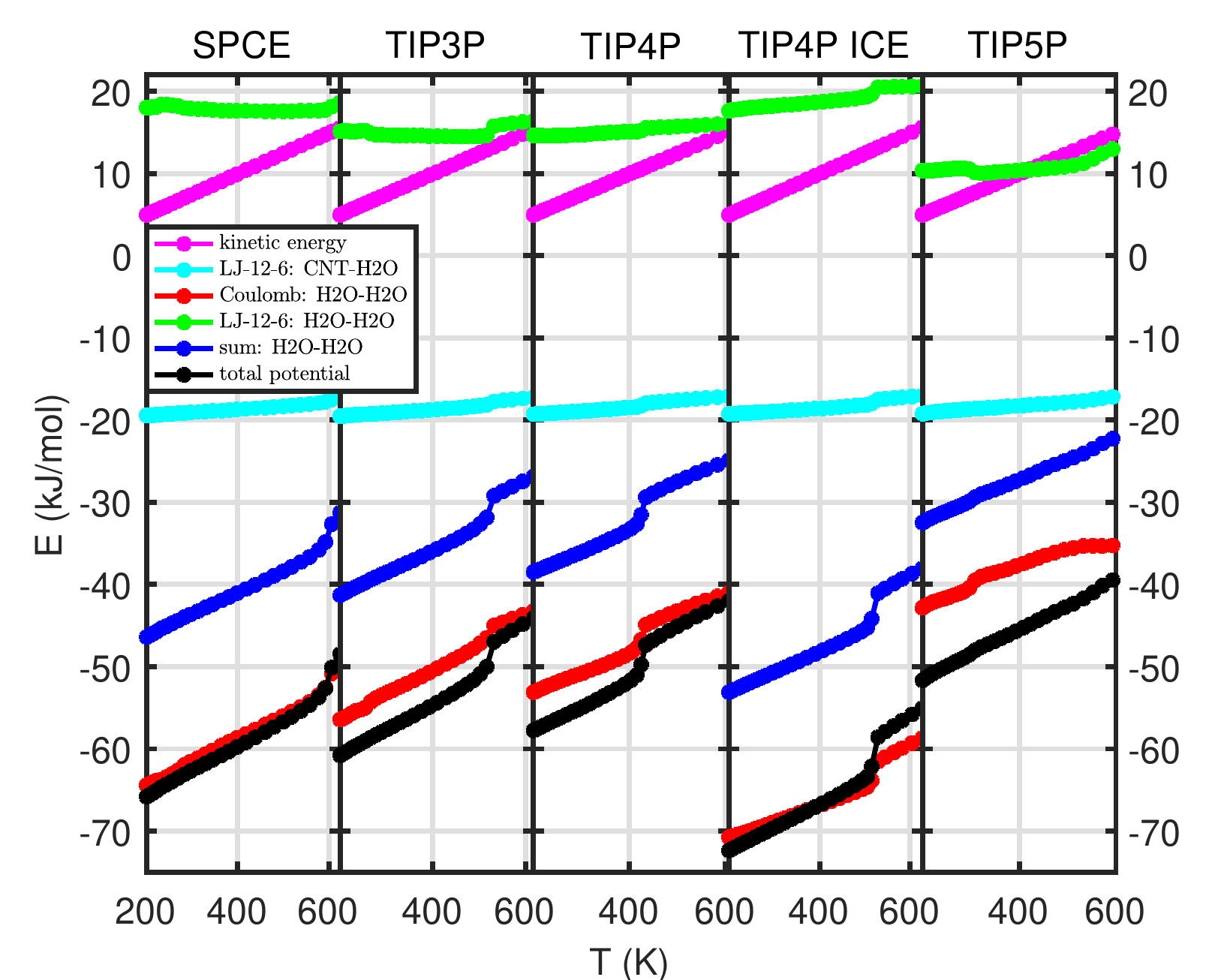}
\caption{Decomposition of total energy in REMD simulations of water confined inside a graphene nanocapillary as a function of temperature. For $\eta=1$ kJ/mol, $\delta=1$, and for SPCE, TIP3P, TIP4P, TIP4P/ICE, and TIP5P water model.}
\label{energy}
\end{figure}

\begin{figure}
\centering          
\includegraphics[width=1.0\textwidth]{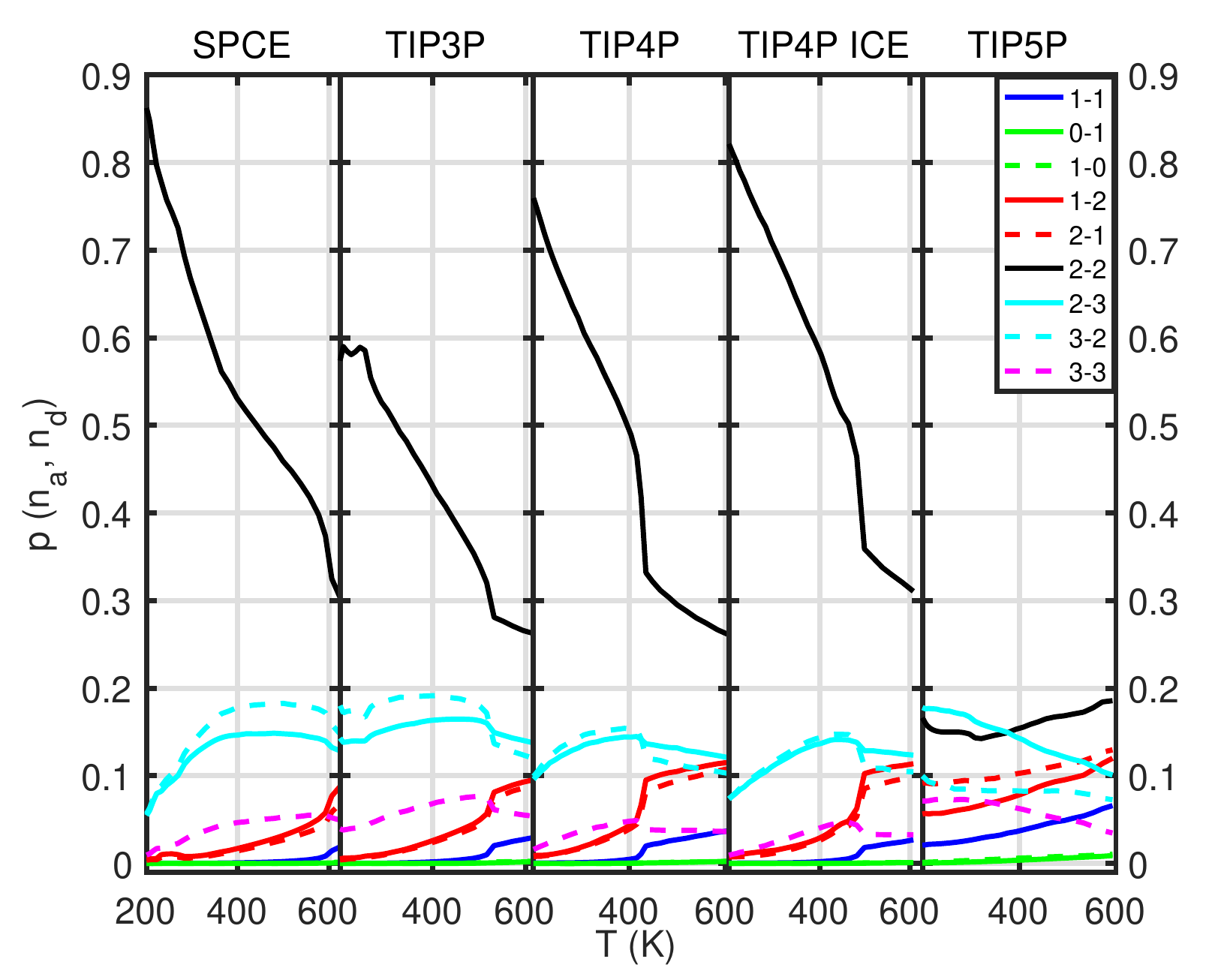}
\caption{H--bonding pattern of water confined inside a graphene nanocapillary as a function of temperature. Obtained from REMD simulations with $\eta=1$ kJ/mol, $\delta=1$, and for SPCE, TIP3P, TIP4P, TIP4P/ICE, and TIP5P water models. 
The curves indicate the joint probabilities, $p_{n_a,n_b}$, of a water molecule acting $n_a$ times as an acceptor and $n_d$ times as a donor, as indicated in the figure legend.}
\label{H-B}
\end{figure}

From $T=200$ K to $T=283$ K, the distribution of peaks in the angle histogram of Fig.~\ref{Angle_surface} indicates that the oxygen atoms are organized in a flat nearly square (FNS) phase as shown in Fig.~\ref{min_E_structures} (a) with peaks around $72^{\circ}$, $90^{\circ}$, $108^{\circ}$, and $160^{\circ}$. 
At $T=283$ K the sharp peaks become blurred indicating that the FNS--like structure starts to undergo a change. 
When the temperature further increases up to $T=296$ K, suddenly two wide angular peaks appear to $72^{\circ}$, $108^{\circ}$ which corresponds to the flat rhombic (FR) phase as shown in Fig.~\ref{min_E_structures} (b). 
Hence, between $T=283$ K and  $T=296$ K  there is a structural transition from phase FNS to phase FR. 
However, the proton ordering shows a more complicated behavior, see the distribution of the polarization $\langle \mu \rangle$ in Fig.~\ref{Mtot}. 
From $T=200$ K to $T=313$ K, the ice phase is essentially AF with a very low polarization value, similar to that of FNS structure a. From $T=329$ K to $T=345$ K, the ice phase is FI with the polarization taking on several intermediate values with  $0.25 \leq \langle \mu \rangle \leq 0.45$, indicating coexistence of different sub-domains of structure a (FNS) and b (FR). 
From $T=364$ K to $T=577$ K, the ice phase is still FI but displaying a higher polarization $\langle \mu \rangle\approx 0.75$, similar to the corresponding value of FR structure b. 
Other quantities of interest such as the caloric curves (Fig.~\ref{energy}) and hydrogen bonding pattern (Fig.~\ref {H-B}), however, are not affected by the above transition, because the phases a (FNS) and b (FR) are nearly iso-energetic and share the same number of hydrogen bonds.  
Hence, the solid--solid structural transition between $T=283$ K and $T=296$ K is classified as a second--order phase transition.

Fig.~\ref{Angle_surface} shows another phase transition between $T=577$ K and $T=596$ K, marked by the disappearance of the peaks in the angular distribution at $72^{\circ}$ and $108^{\circ}$. 
For temperatures above this transition, we observe a loss of the rhombic structure, leading to a liquid-like phase which is also confirmed by the abrupt rise of the self-diffusion constant $D$, see Eq.~(\ref{diffusion}).
In the angular histogram of Fig.~\ref{Angle_surface}, a new peak arises around $60^{\circ}$, mainly caused by nearly triangular configurations appearing in the irregular liquid structure. 
Also between $120^{\circ}$ and $140^{\circ}$ the intensity increases smoothly, indicating the coexistence with irregular tetragons and pentagons.
The detection of this melting-like transition is also supported by the temperature dependence of the hydrogen--bonding patterns shown in Fig.~\ref {H-B}.
At $T=577$ K the probability of fourfold coordination, $p_{2,2}$, starts to decrease drastically while that for three-fold coordination, $p_{1,2}$, starts to rise.
Moreover, Fig.~\ref{Mtot} shows that there is also a substantial loss of proton ordering in the same temperature range, leading from FI to AF arrangements.
In summary, this transition from an ordered quasi-2D crystal to a liquid-like phase is a first-order transition, which is also clear from Fig.~\ref{energy} showing a sudden increase of the total water potential energy by nearly 4.0 kJ/mol (obtained from linear fits to the water--water potential energy below and above the transition temperature). 
This latent heat is higher than the bulk ice melting energy of 3.1 kJ/mol obtained for SPCE water simulations which, however, largely underestimates the experimental value of 6.029 kJ/mol \cite{Curtiss1979a} at 273 K for ambient pressure (0.1 MPa), see also Tab.~\ref {melting enthalpy}. 
Obviously, the confinement of water inside a graphene capillary with a high pressure of 1000 MPa causes the substantial increase of the transition temperature from 215 K to 577 K for SPCE water model. 

\subsubsection{TIP3P Water Model}

At $T=200$ K the distribution of peaks in the angle histograms in Fig.~\ref{Angle_surface} indicates that the oxygen atoms are organized in a flat rhombic (FR) arrangement, similar to that shown in Fig.~\ref{min_E_structures} c, with sharp peaks centered at $77 ^{\circ}$ and $103^{\circ}$. 
However, with the temperature increasing from $T=200$ K to $T=532$ K, the peaks become blurred which indicates that the phase c (FR) is gradually disappearing. 
Within that temperature range, the quasi-2D ice is FI with the polarization $\langle \mu \rangle \approx 0.74$ being practically constant, see Fig.~\ref{Mtot}.
The two quantities show a clear structural transition from the ordered structure c (FR) to a liquid-like phase occurring between $T=532$ K and $T=551$ K. 
As for the SPC/E results discussed above, this melting-like transition can also be detected from characteristic changes of the O angles as well as from the loss of proton ordering.
This transition also shows up in the self-diffusion constant $D$ as well as in the energy decomposition shown in Fig.~\ref{energy} where all curves show an inflection at 532 K with a latent heat around 3.4 kJ/mol which is much higher than the bulk water melting energy of 1.3 kJ/mol from TIP3P simulations.
Again, the melting temperature of the quasi-2D water is much higher than for bulk water, which is due to the confinement and the high pressure of 1000 MPa. 
  
The melting-like transition is also reflected in the structure of the H--bonding networks.
Fig.~\ref {H-B} shows that from $T=200$ K to $T=532$ K the joint probability $p_{2,2}$ decreases continuously indicating that the quasi-2D ice structure disappears. 
Between $T=551$ K and $T=617$ K the joint probability $p_{2,2}$ decreases further (but more slowly) while there are increasing probabilities of defects with two--, three--, and even five--fold coordination. 
 
\subsubsection{TIP4P Water Model}

At $T=200$ K, the distribution of the peaks in the angle histogram of Fig.~\ref{Angle_surface} shows that the oxygen atoms are organized in a nearly square (PNS) fashion, similar to that shown in Fig.~\ref{min_E_structures} (d), with peaks at  $75 ^{\circ}$, $90^{\circ}$, and $105^{\circ}$. 
From $T=200$ K to $T=245$ K, the sharp peak at $90^{\circ}$ becomes blurred and eventually vanishes for $T=256$ K while 
the other peaks at $75^{\circ}$ and $105^{\circ}$ remain essentially unchanged.
This indicates a structural change of the quasi-2D ice from nearly square phase d (PNS) to rhombic phase e (PR) in the temperature range between $T=245$ K and $T=256$ K. 
Moreover, the transition from phase d to phase e is confirmed by the distributions of polarization $\langle \mu \rangle$ in Fig.~\ref{Mtot}.
However, there the transition between the phases (both of which are AF) is found at a higher temperature of $T=359$ K.
The temperature dependence of the energy decomposition, see Fig.~\ref{energy}, and of the hydrogen bonding pattern, see Fig.~\ref {H-B}, however, are less sensitive to this transition because the two phases are nearly isoenergetic and share the same hydrogen bonding pattern. 
Again, this solid--solid structural transition is classified as a second--order phase transition. 
It is worth mentioning that both the square and the rhombic TIP4P quasi-2D phases are puckered which is quite different from the water structures obtained for the SPCE and TIP3P water models. 
 
Both Figs.~\ref{Angle_surface} and \ref{Mtot} show another phase transition between $T=424$ K and $T=434$ K with the sudden disappearance of the peaks in the two types of histograms. 
Again, this is a melting-like transition from a solid to a liquid-like phase, similar to those discussed for the three-site water models.
This first--order transition is also detected in the H-bond patterns shown in Fig.~\ref{H-B} and in the energy decomposition shown in Fig.~\ref{energy}.
There, all curves for water show an inflection at $T=424$ K which is much higher than the melting temperature for TIP4P bulk water at $T=232$ K.
Our value for the quasi-2D water latent heat around 3.3 kJ/mol is below the bulk water melting energy $4.4$ kJ/mol from TIP4P simulations, see Tab.~\ref {melting enthalpy}. 

\subsubsection{TIP4P/ICE Water Model}

The TIP4P/ICE model is a variant of the original TIP4P model, intended to improve the properties of ice and the phase diagram for bulk  water.
In particular, the predicted melting temperature of hexagonal ice (Ih) at 0.1 MPa is now at a much more realistic value of 272.2 K.
The results of our simulations for quasi-2D water using the TIP4P/ICE model can be seen in the angle histograms of Fig.~\ref{Angle_surface}. 
At $T=200$ K, the  peaks at $75 ^{\circ}$, $90^{\circ}$, and $105^{\circ}$ suggest that the oxygen atoms are organized in a nearly square fashion corresponding to phase f (PNS), see Fig.~\ref{min_E_structures} (f).   
From $T=200$ K to $T=515$ K, the three peaks become more and more blurred, thus indicating an increasing flexibility of the quasi-2D ice structure. 
Throughout this temperature range, the polarization $\langle \mu \rangle$ remains very low (see Fig.~\ref{Mtot}), in accordance with the AF character of structure f. 

Between $T=515$ K to $T=529$ K, both the oxygen and hydrogen atoms become disordered and the structure changes from solid to liquid-like. 
The temperature for this melting-like transition is considerably higher than for the original TIP4P model.
Again, this transition is a first--order phase transition which is confirmed the hydrogen bonding pattern (Fig.~\ref {H-B}) as well as the energy decomposition (Fig.~\ref{energy}).
The latent heat of 3.8 kJ/mol is less than the bulk water melting energy 5.4 kJ/mol from TIP4P/ICE simulation, both of which are notably higher than the corresponding values for the TIP4P model, see again Tab.~\ref {melting enthalpy}.

\subsubsection{TIP5P Water Model}

\begin{figure}
\centering
\includegraphics[width=0.7\textwidth]{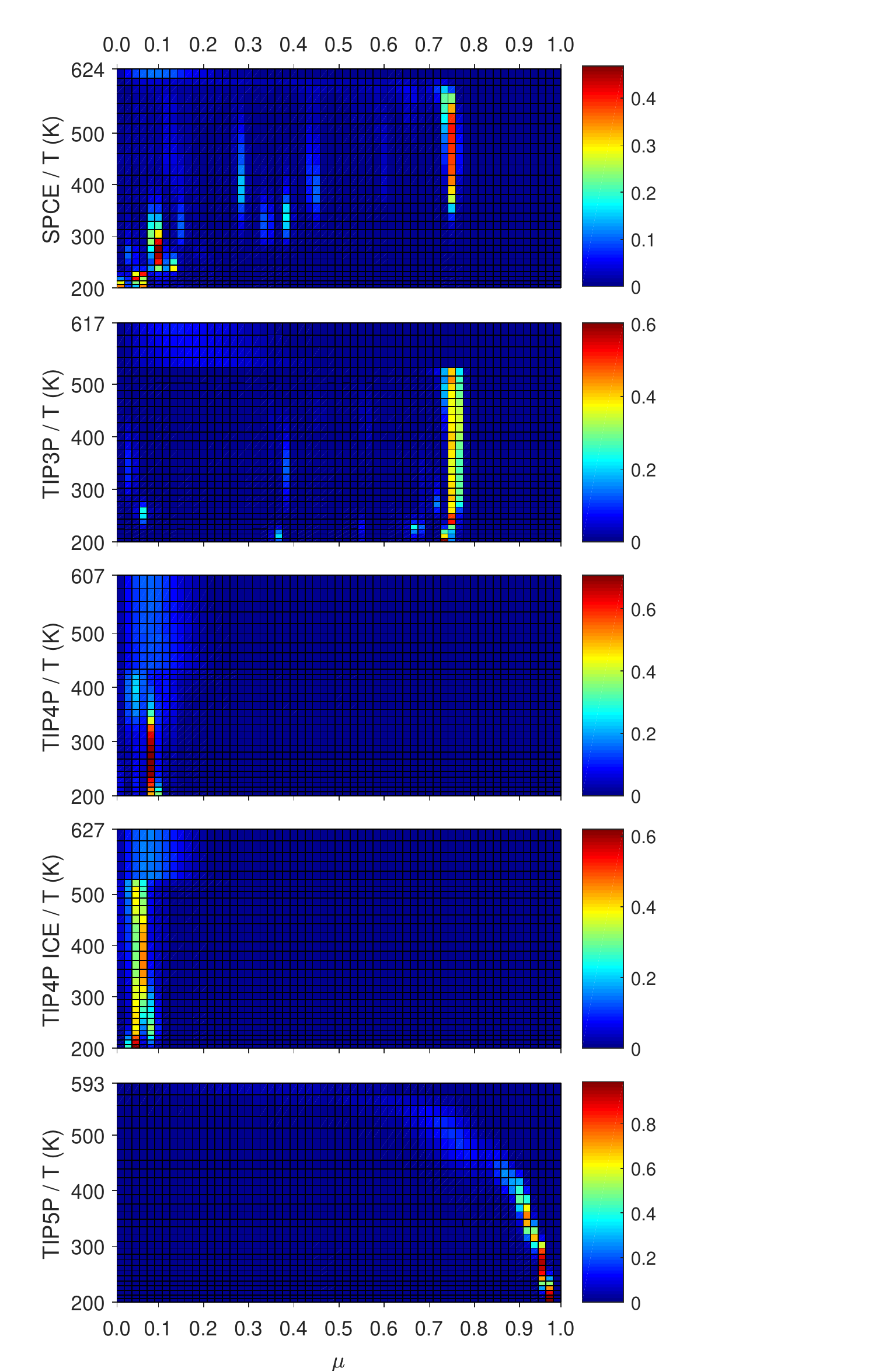}
\caption{Distribution of polarizations $\langle \mu \rangle$ for water confined inside a graphene nanocapillary as a function of temperature. Obtained from REMD simulations with $\eta=1$ kJ/mol, $\delta=1$, and for the SPCE, TIP3P, TIP4P, TIP4P/ICE, and TIP5P water models from top to bottom for a pressure of 1000 MPa.}
\label{Mtot}
\end{figure}

From $T=200$ K to $T=229$ K, the distribution of peaks in the angle histograms in Fig.~\ref{Angle_surface} indicates a flat rhombic oxygen structure.
The peaks are located around $70^{\circ}$ and $110^{\circ}$ resembling those found for structure g (FR) which is also confirmed by the high polarization (FE) of $\langle \mu \rangle \approx 0.97$ in Fig.~\ref{Mtot}.
From $T=238$ K to $T=297$ K, the angular peaks shift and their distribution becomes broader, thus becoming similar to structure h (PR) as shown in Fig.~\ref{min_E_structures}, with the oxygen atoms being more irregular and also weakly puckered. 
This ice phase is also FE with very high polarization value $\langle \mu \rangle \approx 0.95$, but slightly lower than for the lowest temperature phase, as shown in Fig.~\ref{Mtot}.
The transition between the FR and PR phases occuring in the temperature range from $T=229$ K to $T=238$ K is a second-order transition since no latent heat is involved, see the energy decomposition shown in Fig.~\ref{energy}.

Between $T=309$ K and $T=533$ K, the angular distributions in Fig.~\ref{Angle_surface} does not change notably.
The low self--diffusion constant $D$ indicates that the water is still solid--like, but a careful inspection of snapshots of the structures reveals that the ice is amorphous.
In this temperature range, the polarization $\langle \mu \rangle$ is gradually decreasing, corresponding to a transformation from FE to FI proton arrangement, as shown in Fig.~\ref{Mtot}. 
Note that both the LJ and the Coulombic parts of the water--water interaction energy displays a slight inflection upon the transition from ordered to amorphous ice, occurring between temperature $T=297$ K and $T=309$ K.
Because of the different signs of the energy changes of the two contributions, they almost cancel each other, thus rendering also this transition second ordered phase transition. 

When raising the temperature to 553 K, there is a rapid onset of self-diffusion with $D$ rising from values below 0.1 to about 0.7 nm$^2$/ns indicating a melting-like phase transition between $T=533$ K and $T=553$ K.
Interestingly, this transition is hardly visible in the angular distributions of Fig.~\ref{Angle_surface}, 
neither does the polarization shown in Fig.~\ref{Mtot} reflect this transition clearly.
Also the latent heat is negligible; it cannot be compared with the value for bulk water melting (7.3 kJ/mol from simulation and 6.029  kJ/mol from experiments at 273 K and 0.1 MPa pressure) as shown in Tab.~\ref {melting enthalpy}. 
\section{Summary and conclusions}

In the present work we have investigated the structure and dynamics of quasi--2D water confined in between two layers of graphene at a distance of 0.65 nm. 
In a first series of simulations we determined the water occupancy of the nanocapillary  for five different water--water interaction models using MD simulations. 
It is found that the differences in the water occupancy at high pressure (1000 MPa) are negligible.  
Furthermore, neither the total water--carbon interaction strength, $\eta$, nor the corresponding anisotropy parameter, $\delta$, have notable effects on the water occupancy.

Based on the water occupancies obtained from these filling simulations, we studied minimum energy quasi--2D ice structures for the different water models.
The main difference from bulk ice is the complete absence of hexagonal structure; instead only tetragonal arrangements are found.
Depending on the tilt angle, $\tau$, these tetragons are classified as nearly square
or as rhombic.
Both these classes of structures are found for the SPCE and TIP4P water models at very similar energy.
However, only nearly square patterns are observed for water model TIP4P/ICE whereas
in other cases (TIP3P, TIP5P) only rhombic minimum energy structures are detected.
Note the analogy with our previous studies of water confined inside low--diameter carbon nanotubes (CNTs) \cite{Li2015}.
When unrolling the single--walled ice nanotubes (INTs), very similar water networks were found.
For near--zero, medium, or strong tilt of the tetragons, the INT structures could be classified as prism-like, single helix, or double helix, respectively. 

In addition, we studied the effect of using different water models on the proton ordering of quasi-2D ice.
Our calculations revealed antiferroelectric structures for SPCE, TIP4P and TIP4P/ICE, whereas ferrielectric arrangements are found for SPCE and TIP3P.
Only for TIP5P also ferroelectric quasi-2D ice was detected, which is partly different from our previous results for single walled INTs inside CNTs where also various ferrielectric and antiferroelectric structures were found \cite{Li2015}.

In conclusion, by comparing both oxygen arrangements and proton ordering, we showed that the choice of water models plays a key role in determining the outcome of our simulations.
Obviously, this makes a direct comparison with previous work in the literature difficult.
In addition, such comparisons are hampered by different values for the pressure, temperature and the graphene slit width, as well as different assumptions for the water--graphene interaction.
One should keep in mind that our potential energy function, based on high--level electronic structure calculations \cite{Perez-Hernandez2013,Lei2016}, is less hydrophobic and more  anisotropic than in most calculations found in the literature \cite{Algara2015,Kaneko2014,Koga2005,Zhu2015a,Zhao2014a,Bai2010a}. 
The experimental evidence of square monolayer ice is of key importance, see Ref.~\cite{Algara2015} . 
In that work also simulation results were reported for SPCE water, indicating nearly square networks of water molecules, very similar to our findings for the same water model.
Also Ref.~\cite{Zhu2015a} is of particular importance, where it is shown that these  nearly square quasi--2D ice structures for TIP4P water are found only when the pressure exceeds the compression limit of a few hundred MPa, and that these structures are flat only when the slit width is below 0.67 nm.
Such ice structures were also found in other simulation studies of TIP4P water but confined between other hydrophobic surfaces \cite{Kaneko2014}. 
In contrast to our results, nearly square ice was also detected in TIP5P simulations, however, for lower temperature and pressure~\cite{Bai2010a}. 

Rhombic quasi-2D arrangements, which we found for almost all water models investigated, were also predicted in the TIP5P water simulations of Ref.~\cite{Zhao2014a}, however, for lower temperature, lower pressure, and narrower graphene slits.
In addition, such ice structures were also found in other simulation studies based on the TIP4P water model \cite{Koga2005,Kaneko2013} as well as TIP5P water model \cite{Zangi2003b} but for water confined in capillaries of other hydrophobic materials. 
 
Apart from simple tetragonal ice structures, also certain Archimedean tiling patterns were obtained for water adsorbed on a fully hydroxylated silica surface based on density-functional theory (DFT) calculations \cite{Yang2004}.
Subsequently, such patterns were also found in TIP5P simulations of water confined in graphene slits, however, for rather low water density and hundreds of MPa negative lateral pressure \cite{Bai2010a,Zhao2014a}.
 
Note that also hexagonal quasi-2D ice phases were observed in TIP5P water but only for lower water densities, narrower graphene slits, weaker lateral pressures and lower temperatures than in our simulations~\cite{Zhao2014a}.
Similar structures were also predicted for SPCE water confined between hydrophobic plates \cite{Ferguson2012a}.
 
While there is no experimental evidence yet for ferroelectricity of water monolayers inside graphene nanocapillaries, an atomic force microscopy (AFM) imaging work suggests the possibility that water monolayers adsorbed on mica surfaces are ferroelectric \cite{Spagnoli2003a}. 
Subsequently, ferroelectric proton ordering was found in TIP5P MD simulations both for the above-mentioned hexagonal and rhombic monolayer ice~\cite{Zhao2014a}. 
Note that those proton ordered rhombic structures are comparable to our results for the same water model.
In other studies based on the TIP5P water model, however, neither ferroelectric nor ferrielectric water orientations were found~\cite{Bai2010a}, probably due to different pressures applied.

To the best of our knowledge, quasi-2D ferrielectric ice structures for SPCE and TIP3P water models were observed in our work for the first time. 
Our finding of anti-ferroelectric ice structures for four--site water models is in agreement with previous results \cite{Koga2005,Zhu2015a}, despite the different assumptions, i.e. hydrophobic water--graphene force field and isotropic water--graphene interaction.

The resulting temperature-dependence of the structural properties of quasi-2D water reveals intriguing transition phenomena:
In particular, we encountered two classes of phase transitions.
Firstly, there are structural transitions between different solid phases which, in most cases, are similar to some of the minimum energy structures mentioned above. 
Because these structures are normally very close in energy, the transitions between them are classified as second--order transitions, without latent heat involved.
Examples are the FNS--FR transition found for the SPCE water model between $T=283$ K and $ T=296$ K, the PNS--PR transition found for TIP4P between $T=245$ K and $T=256$ K and the FR--PR transition found for TIP5P between $T=229$ K and $T=238$ K. 
Second, there are melting--like transitions between solid and liquid phases of quasi-2D water.
These transitions are classified as first--order transitions, with a notable latent heat (with TIP5P being an exception).
The value of the latent heat, however, is model dependent.
It ranges from 3.3 kJ/mol (TIP4P) to 4.0 kJ/mol (SPCE), which is in all cases different from the respective simulation results for bulk water which in turn are considerably below the experimental value of 6 kJ/mol for bulk water. 
Also the temperatures at which those melting--like transitions occur are strongly dependent on the water model:
The transition temperatures range from 424 K (TIP4P) to 577 K (SPCE). 
All of these temperatures are much higher (factor 1.5 to almost 3) than the corresponding temperatures found in simulations of bulk water using the same water models.
Interestingly, those temperatures are comparable with the phase transition temperatures of water confined inside carbon nanotubes observed using Raman spectroscopy \cite{Agrawal2016}, revealing reversible melting between 378 K and 424 K (360 K and 390 K) for 1.05 nm (1.06 nm) diameter single-walled carbon nanotubes, respectively.

\begin{acknowledgments}
S. L. is grateful to the Chinese Scholarship Council for financial support. The authors would like to thank the HPC Service of ZEDAT, Freie Universit\"{a}t Berlin, for computing time.
\end{acknowledgments}
\bibliography{WaterGraphene}

\clearpage
\newcommand{\tabincell}[2]{\begin{tabular}{@{}#1@{}}#2\end{tabular}}
\begin{table}
\begin{tabular}{cc}
\hline\hline
  water model              & temperatures (K)                     \\                                                                                            \\
\hline
SPCE     & \tabincell {l}{200 207 214 222 232 244 256 269 283 \\296 313 329 345 364 381 398 418 438 \\460 479 499 518 538 557 577 592 606\\ 624}                                                                                                    \\
\hline
TIP3P     & \tabincell {l}  {200 207 215 224 233 243 254 266 277 \\289 302 315 329 343 359 375 392 409\\ 427 443 458 473 488 502 516 532 551 \\571 594 617}                                          \\  
\hline                          
TIP4P     &\tabincell {l}  {200 207 216 224 234 245 256 268 281 \\293 305 318 332 346 359 374 389 403 \\415 424 434 448 464 482 499 518 539 \\559 583 607}                        \\
\hline
TIP4P/ICE     &\tabincell {l}  {200 207 216 225 235 246 258 270 281 \\295 308 322 338 354 371 389 407 425 \\443 460 478 493 505 515 529 545 564 \\582 604 627}    \\
\hline
TIP5P  & \tabincell {l} {200 206 213 221 229 238 247 256 266 \\ 276 286 297 309 322 335 349 362 376 \\393 409 423 439 455 474 493 512 533\\ 553 572 593 }                                     \\      
\hline\hline
\end{tabular}
\caption{Temperature distributions used in REMD simulations for water inside graphene nanocapillaries for different water models SPCE, TIP3P, TIP4P, TIP4P/ICE, and TIP5P. In total, every REMD simulation is 20 ns long.}
\label{Temp}
\end{table}

\clearpage
\begin{table}
\begin{tabular}{clcccr}
\hline\hline
Pressure (MPa)        &SPCE  &TIP3P     &TIP4P   &TIP4P/Ice     &TIP5P\\
\hline
0.1      & 12.98  & 12.93      &11.89     &13.17      &11.49\\
1000    & 13.84  &13.70       &13.59     & 13.83      &13.55 \\                                 
\hline
\end{tabular}
\caption{Density of water confined in graphene nanocapillaries (molecules per $nm^2$) for different water models. For $T=300$ K and for graphene--water interaction parameters $\eta=1$ kJ/mol and $\delta=1$.}
\label{water number}
\end{table}

\clearpage

\begin{table}
\begin{tabular}{c c c c c c c }
\hline\hline
 & structure&  water model   &$\langle\mu\rangle$ &$\alpha$           &$E_{W-W} $   &$E_{W-C}$          \\
\hline
a&   FNS&  SPCE             &0.05    &75  90     105  165                       &  -45.66          &-20.02 \\
b&   FR&  SPCE             &0.75    &77 103  152    179                         &-45.01            &-19.99 \\

\hline
c&   FR&  TIP3P           &0.74    &77  80 100 103 156 179                    &-45.47             &-20.14\\
\hline

d&    PNS&  TIP4P             &0.09  &  75 81 90 99 105  163               &  -43.68           &-19.76\\
e&   PR&  TIP4P           &0.02     &  75 85 95 105 152 168                      &-42.96         &-19.41\\

\hline
f&   PNS& TIP4P/ICE      &0.03   &  75 90 105 156 165                &-57.83           &-19.60\\
\hline
g&   FR&TIP5P          &0.97    &    71 101 116 172                  &-36.90         &-19.98\\
h &    PR&TIP5P          &0.94      &    60 71 82 100 109 118 131 169     &-36.70          &-19.79\\

\hline\hline
\end{tabular}
\caption{Different minimum energy quasi--2D ice structures found for water confined in graphene nanocapillaries simulated with different water models. 
The structures are characterized by their polarizations $\langle\mu\rangle$ (dimensionless) and H-bond networks characterized by angle $\alpha$ (degree). Finally, $E_{W-W}$ (kJ/mol) and $E_{W-C}$ (kJ/mol) denote the water--water and water--carbon potential energies per molecule, respectively. The structures are denoted as follows: F = flat; P = puckered; N = nearly; S= square and R = rhombic.
Note that these structures are also shown in Figs.~\ref{min_E_structures}, \ref{angle}.}
\label{INT_table}
\end{table}

\begin{table}
\begin{tabular}{c c c c c c c c } 
\hline\hline
    & & SPCE & TIP3P & TIP4P & TIP4P/ICE &  TIP5P & Expt.\\
\hline
$T_m$ & Bulk water$^{a}$      & 215 & 145.6 & 232 & 272.2 & 273.9 & 273.15 \\
$T_m$ & Confined water        & 577 & 532   & 424 & 515   & 297   & --     \\
$\Delta E$ & Bulk water$^{a}$ & 3.1 & 1.3   & 4.4 & 5.4   & 7.3   & 6.029  \\
$\Delta E$ & Confined water   & 4.0 & 3.4   & 3.3 & 3.8   & 0.23  & --     \\
\hline
\end{tabular}
\caption{Melting temperatures (K) and latent heats (kJ/mol) of bulk ice and quasi--2D confined ice ($P=1000$ MPa) simulated using different water models versus experimental values. $^a$ From Ref. \cite{Vega2005} (SPCE, TIP3P,TIP4P,TIP5P), Ref. \cite{Abascal2005} (TIP4P/ICE)}
\label{melting enthalpy}
\end{table}

\clearpage

\clearpage

\clearpage

\clearpage

\clearpage

\clearpage

\clearpage

\clearpage

\clearpage

\end{document}